\documentclass[letter]{aa}
\usepackage{natbib}
\bibpunct{(}{)}{;}{a}{}{,} 
\usepackage{setspace}
\usepackage{graphicx}
\usepackage{tabularx}

\begin{document}

\title{Direct imaging
 discovery of 12-14 Jupiter mass object orbiting a young binary system
 of very low-mass stars.
} 
 

\author{
  P. Delorme,  \inst{1} 
  J. Gagn\'e, \inst{2} 
  J.H. Girard, \inst{3}
  A.M. Lagrange, \inst{1} 
  G. Chauvin, \inst{1}
  M-E. Naud, \inst{2}
  D. Lafreni\`ere, \inst{2}
  R. Doyon, \inst{2}
  A. Riedel, \inst{4}
  M. Bonnefoy \inst{5}
  and L. Malo \inst{2}}

\offprints{P. Delorme,
  \email{Philippe.Delorme@obs.ujf-grenoble.fr}. Based on observations
  obtained with NACO on VLT UT-4 at ESO-Paranal
(runs 090.C-0698(A) and 70.D-0444(A). }

\institute{UJF-Grenoble 1 / CNRS-INSU, Institut de Plan\'etologie et d'Astrophysique de Grenoble (IPAG) UMR 5274, Grenoble, F-38041, France.
    \and D\'epartement de physique and Observatoire du Mont M\'egantic,
  Universit\'e de Montr\'eal, C.P. 6128, Succursale Centre-Ville,
  Montr\'eal, QC H3C 3J7, Canada
 \and European Southern Observatory, Alonso de Córdova 3107, Vitacura, Cassilla 19001, Santiago, Chile
  \and Department of Astrophysics, American Museum of Natural History, Central Park West at 79th Street, New York, NY 10034, USA 
  \and Max Planck Institute for Astronomy, K\"onigstuhl 17, D-69117 Heidelberg, Germany. 
}

\abstract{Though only a handful of extrasolar planets have been discovered via
  direct imaging, each of these discoveries had tremendous impact on
  our understanding of planetary formation, stellar formation and
  cool atmosphere physics.}
{Since many of these newly imaged giant planets orbit massive A or
  even B stars we investigated whether giant planets could be found orbiting
  low-mass stars at large separations. }
{We have been conducting an adaptive optic imaging survey to search
  for planetary-mass companions of young M dwarfs of the solar
  neigbourhood, to probe different initial conditions of planetary formation. }
 {We report here the direct imaging discovery of 2MASS J01033563-5515561ABb, a
  12-14~M$_{Jup}$ companion at a projected separation of 84~AU from a
  pair of young 
  late M stars, with which it shares proper motion.
  We  also detected a Keplerian-compatible orbital motion.}
{This young L-type object at planet/brown dwarf mass boundary is the
  first ever imaged around a binary system at a separation compatible with formation in a disc.}

\date{}


\authorrunning{P. Delorme et al.}
\titlerunning{Direct imaging of a 12-14~M$_{Jup}$ object orbiting a M-dwarf
  binary system}
\maketitle

\section{Introduction}
  The discovery of hundreds of extrasolar planets in the last 20 years 
 has radically modified our understanding of planetary
 formation. Though radial velocity and transit detection methods
 have proven by far the most prolific, the few planetary-mass
 companions which have been discovered by 
 direct imaging have
 provided very challenging constraints for formations models,
 especially the core-accretion model \citep{Pollack.1996} that is
 prefered to explain the formation of Solar System planets. 2M1207B,
 discovered by \citet{Chauvin.2004}, with a mass-ratio of 20-25\% is
 too massive with respect to its primary to have formed by core
 accretion, while most of HR8799 \citep{Marois.2008} would be very
 difficult to form in situ by core-accretion. Only $\beta$-Pictoris b
 \citep{Lagrange.2010} fits relatively well with the core-accretion
 scenario. 
Also,
 several imaged substellar companions
 \citep[e.g.][]{Chauvin.2005,Lafreniere.2008,Carson.2012arX} straddle
 the arbitrary -and debated- 13~M$_{Jup}$ planet/brown dwarf
   boundary. For most of these massive planets (or light brown dwarfs) the
 formation mechanism, stellar or planetar, is still debated \citep{Luhman.2006,Bate.2009,Rafikov.2011,Boss.2011,Stamatellos.2011}.\\
 Circumbinary planets, such as  Kepler-16 ABb \citep{Doyle.2011} are
 even rarer and provide peculiar constraints on planetary formation
 scenarios, notably on the influence of binarity on planet-forming
 discs. 

 We present here the discovery of  2MASS J01033563-5515561ABb, hereafter
   2MASS0103(AB)b, a unique  a 12-14~M$_{Jup}$ substellar companion to a late M dwarf
   binary system.

\section{A 12-14M$_{Jup}$ companion orbiting around a young late M
  binary system}
\subsection{Observations and data reduction}
  We imaged 2M0103 in November 2012 (run 090.C-0698(A)), in $L'$
  band as one target of our NACO survey for planetary companions to young
  nearby M dwarfs \citep{Delorme.2012a}. We used NACO infrared wave-front
  sensor and observed in pupil
  tracking (only 12$^o$ of rotation) and cube mode in $L'$, and our
  follow-up observations in $JHK_S$ on the same night used field tracking. Table \ref{obs} shows the details of our observations.

The target star was resolved as a low contrast, 0.25$\arcsec$ binary
on these raw images and an additional source was identified at
$\sim$1.8$\arcsec$ at the north west of 2M0103A. In order to measure the proper
motion of this source, we retrieved ESO archive NACO $H$-band images
of 2M0103, obtained in October 2002 (run 70.D-0444(A)). These early images were acquired in field tracking and
with poor adaptive optics correction. We stacked
the best 50\% of the frames, for which the central binary was resolved,
totalling  100s exposure time on target.\\

\begin{center}
\begin{figure}
\includegraphics[width=9.0cm]{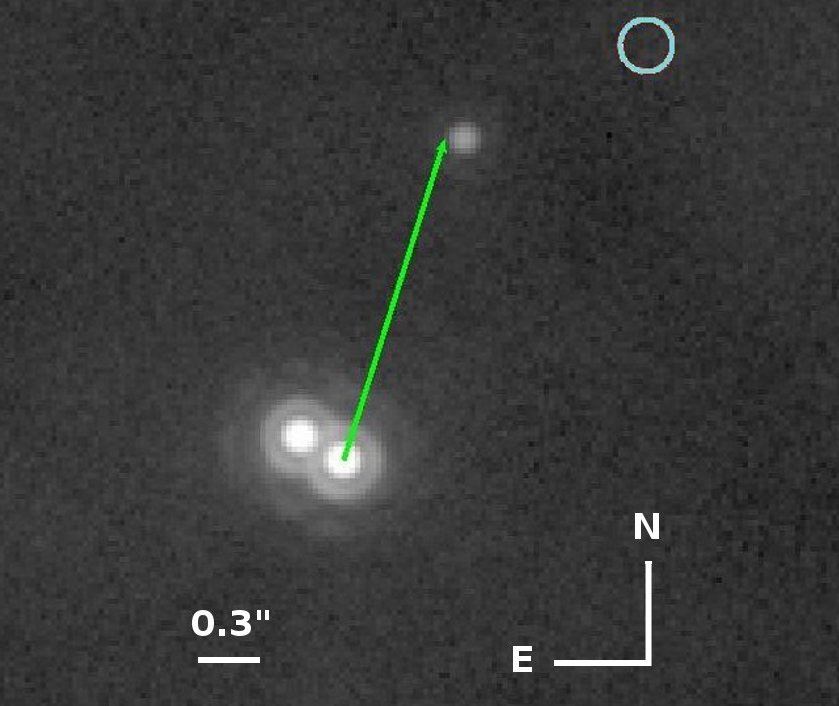}
\caption{\label{PM} 
 2MASS0103(AB)b in November 2012, with NACO in
  $L'$ band. The green arrow shows the position
  of the companion in 2002. The light-blue circle identifies
the expected position of the companion if it had been a background
source.  }
\end{figure}
\end{center}

We used the IPAG-ADI pipeline as described in \citet{Delorme.2012a} to
reduce the frames (bad pixel interpolation, flat, recentring,
derotation and stacking). Although both the
secondary component and the companion appear clearly after a simple
stack of all exposures (see Fig.\ref{PM}), we performed ADI
\citep{Marois.2006} and LOCI \citep{Lafreniere.2007} star subtraction
procedures to detect eventual other companions. None was detected was
detected down to a 
contrast of $\sim$7.5 magnitudes at 0.5$\arcsec$, resulting in a
detection limit of $\sim$2.5~M$_{Jup}$ at 25~AU for an age of 30~Myr
(see discussion below).

\begin{table}
\caption{Summary of the NACO (VLT-UT4) observations of 2M0103AB (RA=01:03:35.63; Dec=-55:15:56.1).\label{obs}}
\begin{tabular}{ccc|c}
UT Date & Filter & Exp. time & Comments \\ \hline
           &  $L'$ & 32$\times$200$\times$0.2=1280s & Seeing:\\
2012-11-25 & $K_S$ & 8$\times$20=160s & 0.7$\arcsec$-0.8$\arcsec$ \\
 & $H$ & 8$\times$20=160s &  Airmass: \\
 & $J$ & 4$\times$5=20s &1.16-1.25\\ \hline
2002-10-28 & $H$ & 5$\times$10$\times$2=100s & Archive data\\ 
\end{tabular}
\end{table}

\subsection{Host star properties}
The primary star 2MASS J01033563-5515561 was identified as part of a
survey designed to identify new, later than M5 candidate stars and
brown dwarfs to the young, nearby moving groups and associations Beta
Pictoris, TW Hydrae, Tucana-Horologium (THA), Columba, Carina,
Argus and AB Doradus (ABDMG) \citep{Torres.2008}. The details of this analysis will be
presented in  Gagné et al. (in prep.), but the principle is to
identify promising candidate members 
to these moving groups using astrometry, proper motion and photometry
from a correlation of 2MASS and WISE catalogs, with a modified version of the
Bayesian analysis described in \citet{Malo.2013}.  One of the first robust
candidates identified in this project is 2MASS J01033563-5515561,
which we have followed with GMOS-S 
at Gemini South to obtain the optical spectra. This spectrum
matches a M5.5/M6 spectral-type and shows strong H-alpha
emission at 656 nm, with an equivalent width of
10.23$\pm$0.55$\AA$. No nearby X-ray source was found in the
ROSAT archive \citep{Voges.1999},
indicating the target is not a strong X-ray emitter.
 In parallel to this, we have obtained a trigonometric distance, of
 47.2$\pm$3.1~pc for this object  (A. Riedel, private communication,
 using the CTIO 0.9m through the CTIOPI program, using 49 $R$-band
 images taken on 11 nights between  October 26th, 2007 and November
 13th, 2012, and reduced using methods from \citet{Jao.2005,Riedel.2011}. The complete parallax analysis for 2M0103, together with many
 other objects, will be published in Riedel et al., in preparation. During the NACO runs described earlier, we have also noticed
 the primary is in fact a binary with a flux ratio of 0.8 in the
 $L'$ band. Taking into account this binarity and the trigonometric
 distance, we find Bayesian probabilities of 99.6\% and  0.4\%
 for  membership to THA and ABDMG respectively. The field hypothesis
 has a probability of 10$^{-14}$. 2M0103AB is therefore a strong candidate member of the Tucana-Horologium
 association, aged $\sim$30~Myr \citep{Torres.2008}. \\
We must stress that those probabilities are not absolute ones in the
sense that even a sample of candidates with a 100\% Bayesian
probability will contain a certain number of
false-positives. Follow-up observations of robust candidates in \citet{Malo.2013}  have shown that the false-positive rate is 10\% for
candidates without parallax in THA. Though the membership analysis in our
study is not exactly the same, the risk of a false positive is very
low, especially because we do have a parallax measurement, meaning
that 2M0103AB is very probably a bona-fide member of THA  We will
assume in the following that the 
 2M0103 system is aged 30~Myr.\\

 According to BT-Settl 2012 isochrones
 \citep{Allard.2012,Baraffe.2003}, and assuming a 
 distance of 47.2$\pm$3.1~pc and an age of 30~Myr, 2M0103AB is a low
 mass binary with masses of [0.19;0.17]$\pm$0.02~M$_{\odot}$ for [A;B]
 respectively, see table \ref{sysmagabs}. The projected separation
 between A and B was 0.26$\pm$0.01$\arcsec$ in 2002 
and 0.249$\pm$0.003$\arcsec$ in 2012. 
The projected distance was around 12~AU at both epochs, but the
position angle changed significantly, from 71.2$^o$
in 2002 to 61.0$^o$ in 2012

\footnotesize
 \begin{table}
\caption{Host system absolute magnitudes compared with BT-Settl
  isochrones at 30Myr
  absolute magnitudes (2MASS for $JHK$ and NACO for $L'$).\label{sysmagabs}}
\begin{tabular}{lcccc}
Filter & $M_J$ & $M_H$ & $M_{K_s}$ & $M_{L'}$ \\ \hline
2M0103A & 7.36$\pm$0.05  & 6.78$\pm$0.05  &  6.44$\pm$0.05 & 6.04$^*$ \\
2M0103B & 7.56$\pm$0.05  & 6.98$\pm$0.05  &  6.64$\pm$0.05 & 6.24$^*$ \\ 
Model~0.2M$_{\odot}$& 7.31  & 6.75  & 6.50  & 6.1\\
~ ~-~ 0.175M$_{\odot}$& 7.51  & 6.95  & 6.70  & 6.3\\
\end{tabular}
\tablefoot{$^*$ Since no calibrated photometry is available in $L'$,
  these magnitudes are derived from the modelled $K_S-L'$.}
\end{table}
\normalsize

\subsection{Proper motion analysis: a bound companion}

During our November 25th, 2012, $L'$ band NACO observations of
2M0103 (run 090.C-0698(A)), we identified a candidate companion with a
separation of 1.78$\arcsec$ and a position angle of 339.3$^o$ from the
primary 2M0103A. Even if contamination by background objects is
relatively low in $L'$ band compared to shorter wavelength \citep[see][]{Delorme.2012a}, the probability
that this companion was a contaminant was high. However, the companion
was redder in $K_S-L'$ than its late M host system,  meaning that the
companion is of even later spectral type, considerably decreasing
the likelyhood of the contaminant hypothesis, but not incontrovertibly
proving  
that the companion is bound. A definite proof of companionship was
however provided thanks to archive images taken with NACO on October
28th, 2002. Using this 10 years time base, we could
determine that 2MASS0103(AB)b unambiguously (contamination probability
$<$0.001\%, taking into account parallax motion) shares the proper 
motion of 2M0103AB (see Fig.\ref{PM}) and is therefore a bound
companion. The relative astrometry at each epoch is shown on table
\ref{astrom}.

\begin{table}
\caption{Separation (Sep.) and position angle (PA) of the companion,
  with respect to 2M0103A and to the center of mass of the binary. \label{astrom}}
\begin{tabular}{l|cc}
  & 2002-10-28   & 2012-11-25    \\ \hline
Sep. from 2M0103A($\arcsec$)     & 1.682$\pm$0.015           &1.784$\pm$0.003  \\
PA from 2M0103A($^o$)            &  341.7$\pm$0.05$^*$    & 339.8$\pm$0.01$^*$  \\ \hline
Sep. from barycenter($\arcsec$) &    1.718$\pm$0.015         &  1.770$\pm$0.003   \\
PA from barycenter($\arcsec$)     &   338.0$\pm$0.05$^*$       &
336.1$\pm$0.01$^*$ \\ \hline
 \end{tabular}
\tablefoot{ $^*$ The error in position angle refer to the relative error between both epochs. The absolute error, dominated by
  systematic uncertainties in the position of the reference stars in theta Ori.,
is $\pm$0.4$^o$.} 
\end{table}

\subsection{Companion properties}
\begin{center}
\begin{figure}
\includegraphics[width=9.0cm]{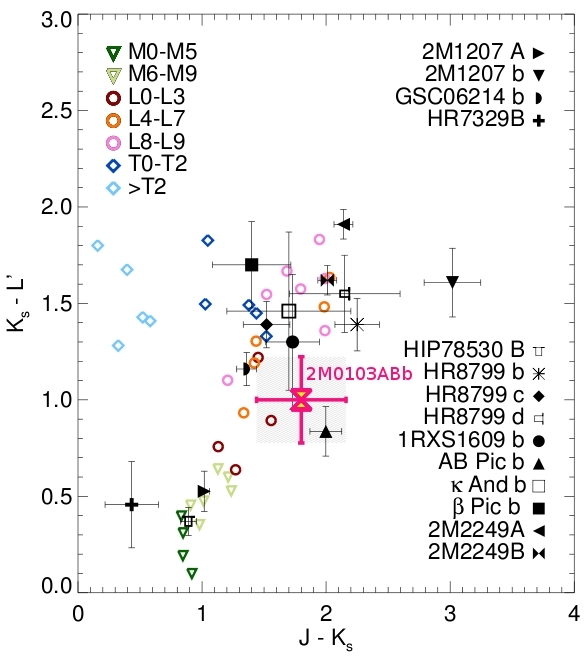}
\caption{\label{colcol}
 $J-K_S$ versus $K_S-L$ colour-colour diagram showing 2MASS0103(AB)b
  together with other known planetary and brown dwarf companions to
  young stars \citep[taken from][]{Bonnefoy.2013arXiv,Bailey.2013arXiv}. The symbols without error bars show the colour of field
M, L and T dwarfs \citep[taken from][]{Golimowski.2004}.}
\end{figure}
\end{center}

Our current information about the physical
   properties of the system relies on the $J,H,K$ and $L'$
   photometry and astrometry from our November 2012 NACO run, as well
   as October 2002 NACO archive images. The resulting Moffat-fitting photometry
   is $J$=15.4$\pm$0.3, $H$=14.2$\pm$0.2, $K_S$=13.6$\pm$0.2 and $L'$= 12.6$\pm$0.1. We emphasize that
   all these measurements were derived relative to 2M0103A,
   which introduces significant systematic errors (hence the large
   errors bars). The absolute  magnitudes of the companion, for a 
   distance of 47.2~pc, are shown in table \ref{compmagabs}, and are
   compatible with 2MASS0103(AB)b being a 12 to 14~M$_{Jup}$ companion orbiting at 84~AU around the young low-mass binary
   2M0103AB. Note that 2MASS0103(AB)b very red colours in $JHKs$ do not
   match the colours of field objects of similar absolute magnitudes
   and are similar to known, early L-type, young planetary-mass objects,
   thus independantly confirming the youth of the system (see also
  Fig.~\ref{colcol}).\\
 It is to be noted that the age of THA is not perfectly known and the dispersion of the age estimations of individual stars in THA
     span the 20-50Myr range \citep{Zuckerman.2000,Torres.2000}. If we
     assume an age of 20 Myr, 2MASS0103(AB)b
   would be a 12-13~M$_{Jup}$ planet, while it would be a 14-15~M$_{Jup}$ brown
   dwarf if we assume an age of 50~Myr. In spite of the naming change,
   the physical differences in mass
   estimates for the 20~Myr and the 50~Myr hypothesis are much smaller
   than those 
   derived in \citet{Marois.2010} for HR8799bcde planets in the same age
   range. An explanation is that objects more massive than $\sim$10~M$_{Jup}$
   undergo some deuterium burning in this age range, somewhat
   compensating cooling down mechanisms. However, since there is
   currently no robust independent mass constraint for any imaged
   exoplanet \citep[saved to some extent for $\beta$Pic.~b, see][]{Lagrange.2012}, it is probable that the systematic uncertainties coming
   from substellar models inaccuracies are larger than those arising
   from age uncertainties. For the sake of comparison with other
   substellar companions found in associations of the same mean age of
   30~Myr, we assume in the following that 2MASS0103(AB)b is a 12-14~M$_{Jup}$ object
   aged 30~Myr.\\
 A possible
   analog, if confirmed as bound, would be the substellar
   object located at a projected separation of 1100~AU from the binary
   system SR12AB \citep{Kuzuhara.2011}. The properties of 2MASS0103(AB)b
   (mass of 12-14~M$_{Jup}$, age of 30 Myr, colours, projected separation of $>$50~AU) and observed colours are also much like  
   AB pic~b \citep[K1V][]{Chauvin.2005} or $\kappa$ Andromeda b \citep[][]{Carson.2012arX}. The properties of the host
   systems are however quite 
   different. While  $\kappa$ Andromeda is a massive B star
   ($\sim$2.5M$_{\odot}$, mass ratio of $\sim$0.5\%),
   2M0103AB is a close binary system composed of 2 late M dwarfs,
   whose combined mass is $\sim$0.36M$_{\odot}$, resulting in a mass
   ratio of approximatively 3.6\% for the system.  \\

The position of the companion at each epoch was derived by
Moffat-fitting and the orientation of the detector was calibrated
using NACO calibration images of theta Ori. obtained close in time of
the science images. As shown on table \ref{astrom}, the relative
astrometry is accurate enough to detect the orbital motion of the
companion around the center of mass of the system, with a projected
motion of 77$\pm$15~mas over ten years. The corresponding velocity at
47.2~pc is 1.7$\pm$0.3~km.s$^{-1}$. The Keplerian
velocity, assuming a
circular orbit of 84~AU around the 0.36M$_{\odot}$ system is
1.96~km.s$^{-1}$, corresponding to a period of 1280 years and is
fully compatible 
with our measurement. 
 It is to be noted that the secondary and the companion rotate in the
same direction and that their observed orbital motion can be
compatible with a face-on orbit but not with an edge-on one. 
\begin{table}
\caption{Companion absolute magnitudes compared with BT-Settl
  isochrones at 30Myr and 5Gyr (field hypothesis)
  absolute magnitudes, and other known companions at the planet/brown
  dwarf mass boundary. \label{compmagabs}}
\begin{tabular}{l|cccc}
Companions & $M_J$ & $M_H$ & $M_{Ks}$ & $M_{L'}$ \\ \hline
2M0103ABb  & 12.1$\pm$0.3  & 10.9$\pm$0.2  &  10.3$\pm$0.2 & 9.3$\pm$0.1\\
$\kappa$And.b  & 12.7$\pm$0.3  & 11.7$\pm$0.2  &  11.0$\pm$0.4 &  9.5$\pm$0.1 \\
ABpic.b  &  12.9$\pm$0.1  & 11.4$\pm$0.1  & 10.8$\pm$0.1  & 9.9$\pm$0.1 \\ \hline \hline 
Models &  &   &  & \\
12~M$_{Jup}$& 12.8  & 11.9  & 11.3  & 10.0\\
15~M$_{Jup}$& 10.9  & 10.3  & 9.9  & 9.0\\
Field&  10.8 &  10.3 & 10.0   & 9.4\\
\end{tabular}
\end{table}

\begin{figure}
\includegraphics[width=8.5cm]{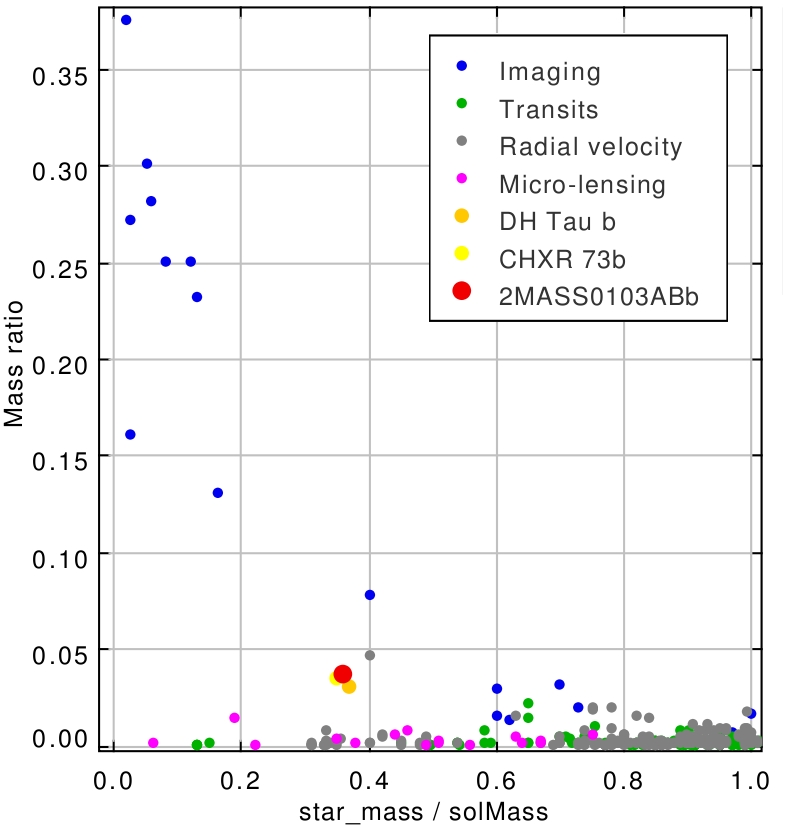}
\caption{\label{mass_ratio} Substellar companion to stellar host mass
  ratio versus stellar host mass, sorted by their discovery technic,
  from the exoplanet.eu database \citep{Schneider.2011}}.
\end{figure}

\section{A challenge for stellar and planetary formation theories}
   2MASS0103(AB)b has a companion mass to host system mass ratio of
   $\sim$0.036, which is too low to match known low 
   mass multiple systems (See Fig.\ref{mass_ratio} and also Allers et
   al., 2007), but still higher than most star-planet systems
   confirmed so far.  Systems with similar mass
     ratio, but almost one order of magnitude more massive,
     with 55-95~M$_{Jup}$  brown dwarfs or very low-mass stars
     orbiting massive (1.2-2M$\odot$) stars, have been
     identified by \citet{Janson.2012}. In case formation mechanisms
     would simply scale with mass, the same processes could be at work
   to explain the formation of these very different kind of objects.
    If we keep to more similar low-mass star systems shown on Fig.~\ref{mass_ratio}, the mass ratio
    of 2MASS0103(AB)b
   is very close to those of DH tau B (8-22~M$_{Jup}$, separation
   of 330~AU) and CHXR 73b (7-20~M$_{Jup}$, separation
   of 210~AU)
   \citep{Itoh.2005,Luhman.2006}, but its projected separation is much
   smaller. \citet{Luhman.2006} state that neither DH tau B nor CHXR
   73b could be formed in situ by core-accretion or disc instability
   because of the very large separation from their host stars, and the
   same holds for the 1100~AU candidate companion to SR12AB   \citep{Kuzuhara.2011}.  The
   case is different for 2MASS0103(AB)b, at a separation of only 84~AU. At
   such separations, a formation in a 
   gravitationally instable primordial circumbinary disk would be
   fully compatible with planetary formation by gravitational instabilities,
   as described by \citet{Boss.2011}. However, this scenario is
   discussed: \citet{Dodson-Robinson.2009} 
   claim that objects formed by disc instabilities around M-dwarfs
   should have $\sim$10\% of the mass of the host system meaning that 2MASS0103(AB)b
   would not be massive enough for such a scenario, while other studies
\citep{Rafikov.2009,Stamatellos.2011} find that such low-mass discs
cannot fragment at all.     
 Simultaneous formation and ejection of the 3 components in the
 massive disc of a more massive orginal host star is 
   plausible, in a scenario akin to what is described in
   \citet{Stamatellos.2009}, but the central binary components, with
   masses of 0.17 and 0.19~M$_{\odot}$ are more massive than most objects
   formed in \citet{Stamatellos.2009} simulations.\\
 A planetary formation 
   scenario by core-accretion
   \citep[e.g.][]{Kennedy.2008,Mordasini.2009,Rafikov.2011} can very probably be
   excluded for several reasons. First, the separation is too large
   for a formation in situ. Second, the
   companion has $\sim$3.6\% of the mass of its host system, which is
   of the order of magnitude of the maximum total mass of the protoplanetary
   disc from which core-accretion planets are formed. Finally, such a
   12-14~M$_{Jup}$ companion would be a very rare
   occurence, according to the
   core-accretion planetary mass function derived by
   \citet{Mordasini.2012}.

 A purely stellar
   formation mode by 
   turbulent core fragmentation \citep[see
     e.g.][]{Padoan.2002,Bate.2009,Hennebelle.2011} is plausible,
   and in this case 2MASS0103(AB)b would be an extreme case
   of hierarchical triple stellar with a third component in the
   12-14~M$_{Jup}$ mass range. However, a stellar formation scenario would
   necessitate that cores can naturally fragment into such low mass
   objects, without requiring any ejection from the the accretion
   reservoir \citep[such as described in][]{Reipurth.2001,Bate.2005}, because it would be
   difficult to starve the accretion of the third component without also
   stopping accretion on the central binary. From hydrodynamical
   simulations of stellar formation by cloud fragmentation, 
   \citet{Bate.2012} claims that ``brown dwarfs with masses  $<$15~M$_{Jup}$
   should be very rare", implying that formation by direct core
   fragmentation of a 
   12-14~M$_{Jup}$ object such as 2MASS0103(AB)b would be possible but uncommon. \\

In any case, the discovery of 2MASS0103(AB)b  brings most current stellar and
planetary formation theories to their limits while others, such as
core-accretion, can probably be excluded. The very existence of such a
peculiar system therefore provides a very valuable test case against
which current and future stellar and planetary formation theoretical
models can be tested.


\begin{acknowledgements}
We acknowledge support from the French National Research Agency (ANR)
through the GuEPARD project grant ANR10-BLANC0504-01. We acknowledge financial support from ''Programme National de Physique Stellaire" (PNPS) of CNRS/INSU, France
\end{acknowledgements}

\bibliographystyle{aa}
\bibliography{biball}

\end{document}